\documentclass[english]{article}
\usepackage[T1]{fontenc}
\usepackage[latin9]{inputenc}
\usepackage{geometry}
\usepackage{amstext}
\usepackage{amssymb}
\usepackage{esint}

\makeatletter

\newcommand*\LyXZeroWidthSpace{\hspace{0pt}}

\usepackage{booktabs}
\usepackage{multirow}
\usepackage{subcaption}
\date{}

\makeatother

\usepackage{babel}
\begin{document}

\title{Deep Hedging Bermudan Swaptions}

\author{Kenjiro Oya\thanks{Nomura Securities Co., Ltd.}}
\maketitle
\begin{abstract}
This paper proposes a novel approach to Bermudan swaption hedging
by applying the deep hedging framework to address limitations of traditional
arbitrage-free methods. Conventional methods assume ideal conditions,
such as zero transaction costs, perfect liquidity, and continuous-time
hedging, which often differ from real market environments. This discrepancy
can lead to residual profit and loss (P\&L), resulting in two primary
issues. First, residual P\&L may prevent achieving the initial model
price, especially with improper parameter settings, potentially causing
a negative P\&L trend and significant financial impacts. Second, controlling
the distribution of residual P\&L to mitigate downside risk is challenging,
as hedged positions may become curve gamma-short, making them vulnerable
to large interest rate movements. The deep hedging approach enables
flexible selection of convex risk measures and hedge strategies, allowing
for improved residual P\&L management. This study also addresses challenges
in applying the deep hedging approach to Bermudan swaptions, such
as efficient arbitrage-free market scenario generation and managing
early exercise conditions. Additionally, we introduce a unique \textquotedbl{}Option
Spread Hedge\textquotedbl{} strategy, which allows for robust hedging
and provides intuitive interpretability. Numerical analysis results
demonstrate the effectiveness of our approach. 
\end{abstract}

\section{Introduction}

\subsection{Motivation for Applying the Deep Hedging Approach to Bermudan Swaptions}

Bermudan Swaptions are a popular type of interest rate derivative.
However, existing methods of managing these positions using arbitrage-free
theory have notable limitations. This paper seeks to address these
limitations by applying the deep hedging approach. 

The main issues with arbitrage-free theory are its unrealistic assumptions,
including zero transaction costs, continuous-time hedging, and unlimited
market liquidity. In reality, financial markets do not operate under
such idealized conditions, and market dynamics often differ from the
theoretical models. As a result, residual profit and loss (P\&L) can
accumulate in real-world trading, leading to the two main issues.

The first issue is the risk of not achieving the initial model price
due to this residual P\&L. This problem is particularly pronounced
when the model parameters are improperly set, which can lead to a
negative P\&L trend with significant financial impact. Unfortunately,
arbitrage-free models offer limited guidance on selecting model parameters.
While they can offer break-even levels at specific times \cite{SMBM},
they lack a systematic way to prevent negative P\&L trends over the
long term.

The second issue is the challenge of controlling the distribution
of residual P\&L to limit downside risk. For example, after hedging
a Bermudan swaption with swaptions using arbitrage-free theory, the
position may become curve gamma-short, making it vulnerable to large
interest rate movements. Even if traders want to mitigate this risk,
it remains unclear how to adjust model assumptions to create a more
favorable gamma-long position\footnote{\cite{SMBM} provides a specific example where performing a hedge
with swaptions results in a curve gamma-short position.}. This complexity arises because the hedge size is linked to model
price sensitivity to market changes, making it hard to predict and
control.

Recent advances in machine learning have opened up new possibilities
in derivative risk management (e.g., \cite{DeepHedgeOriginal,HullExotic}).
In this paper, we use the deep hedging approach \cite{DeepHedgeOriginal},
which enables more flexible risk management by allowing customization
of risk measures and hedging strategies. By adjusting the parameters
of the risk measure function, specifically Conditional Value at Risk
(CVaR), it becomes possible to monitor and manage the distribution
of P\&L drawdowns from the initial model price. This provides a structured
approach to handling P\&L drawdown risk. Furthermore, as CVaR parameters
change, so does the risk profile of the hedged position, allowing
unwanted downside risks to be mitigated.

Furthermore, we explore a new method for mitigating downside risks
by applying constraints to the weights of assets in the hedge portfolio,
preventing over-hedging with the swaption. By combining this method
with CVaR parameter adjustments, we create a more flexible and effective
approach to position management.

\subsection{Key Considerations for Applying Deep Hedging to Bermudan Swaptions}

Applying the deep hedging approach to Bermudan swaptions presents
several technical challenges that must be addressed. These challenges
include: (1) generating training scenarios, and (2) managing early
exercise conditions. Below, we summarize the approaches used to tackle
these challenges.

\subsubsection*{Market Scenario Generation for Training}

First, let us clarify the requirements for training scenarios. As
discussed in \cite{DHRemoveDrift}, when applying the deep hedging
approach to exotic derivatives using options as hedge assets, it is
crucial to ensure that scenarios are both static and statistically
arbitrage-free to achieve stable training results. Additionally, the
scenario generator must be flexible enough to capture the realistic
dynamics of the hedge assets and computationally efficient. In practical
applications, the deep hedging approach requires recalculating model
values and hedged P\&L distributions daily based on current market
conditions, requiring the daily regeneration of scenarios. Therefore,
a scenario generation process capable of operating within realistic
time constraints is essential for practical use.

To meet these requirements, this paper proposes using the Swap Market
Bergomi Model (SMBM) for scenario generation \cite{SMBM}. The SMBM
ensures that scenarios are both static and statistically arbitrage-free,
satisfying the first requirement. Furthermore, the SMBM provides flexibility
in modeling the dynamics of interest rate and swaption processes in
a practical way. The model includes the same number of state variables
as the number of hedge assets, allowing users to define correlation
coefficients between these variables. Additionally, the function that
calculates hedge asset prices from the state variables is a low-dimensional
function, making it suitable for pre-training methods. This enables
scenario generation that is both fast and accurate.

\subsubsection*{Early Exercise Conditions}

For Bermudan swaptions, where the holder has the right to exercise
early\footnote{This paper limits the discussion to position management from the perspective
of the Bermudan swaption holder.}, it is essential not only to model the hedging strategy but also
to accurately represent the early exercise strategy. In this paper,
we propose a recursive model training strategy to train the early
exercise strategy in a stable manner.

Specifically, for a Bermudan swaption with $N$ possible exercise
dates, the training process is broken down recursively. Instead of
training hedging and early exercise strategies for all $N$ exercise
dates at once, the model focuses on each exercise right individually.
The process begins by building a hedging model for a Bermudan swaption
with only the final exercise right. Once this model is trained, it
forms the foundation for developing a model that considers both the
second-to-last and final exercise rights. This recursive approach
is repeated for each earlier exercise right. By isolating the early
exercise strategy for each date, the training process remains stable
and manageable.

In a related study, the application of machine learning to American-style
options has been explored in {[}6{]}, where continuation values are
calculated using deep learning regression based on arbitrage-free
prices, and exercise decisions are made accordingly. Since our aim
is to evaluate risk-adjusted prices, we propose a method that incorporates
exercise decision model training within the deep hedging framework.

\subsection{Structure of the Paper and Our Contributions}

The structure of this paper is as follows:
\begin{itemize}
\item \textbf{Section 2} provides a detailed explanation of the market scenario
generation process. A unique contribution of this research is our
approach to creating arbitrage-free scenarios for Bermudan swaptions
using the Swap Market Bergomi Model combined with pre-training techniques.
\item \textbf{Section 3} describes the methodology for applying the deep
hedging approach to Bermudan swaptions, with particular attention
to handling early exercise features. In this chapter, we introduce
a decomposition technique for Bermudan swaptions and iteratively construct
the hedging model. Additionally, we present a robust hedging strategy
that utilizes swaption spreads as hedging assets. To our knowledge,
this decomposition training approach for Bermudan swaptions and the
robust hedging method are both original contributions.
\item \textbf{Section 4} presents the results of the numerical analysis
based on the proposed approach, providing insights into strategy effectiveness.
\item \textbf{Section 5} concludes the paper, summarizing the findings and
discussing potential directions for future research.
\end{itemize}

\section{Market Scenario Generation}

\subsection{Bermudan Swaptions and Hedge Assets}

The goal of scenario generation for Bermudan swaption hedging is to
construct a joint distribution for interest rate swaps and European
swaptions (referred to as \textquotedbl{}swaptions\textquotedbl{}
from here on). First, we define the hedge assets and Bermudan swaption
prices using equations. Asset prices are used to represent the assets
themselves where there is no ambiguity.

Let $U_{t}^{i,j,K}$ represent the price of a receivers interest rate
swap at time $t$, which starts at $T_{i}$ and matures at $T_{j}$
with a strike of $K$. The following assumes that an interest rate
swap will always refer to a receivers swap. The swap value can be
expressed using the discount factor $P_{t}^{i}$ at time $t$ for
maturity $T_{i}$ as follows:
\begin{eqnarray}
U_{t}^{i,j,K} & = & A_{t}^{i,j}(K-S_{t}^{i,j})\\
A_{t}^{i,j} & \triangleq & \sum_{u=i+1}^{j}\delta_{u-1}P_{t}^{u}\\
S_{t}^{i,j} & \triangleq & \frac{P_{t}^{i}-P_{t}^{j}}{A_{t}^{i,j}}
\end{eqnarray}

Here we use discrete time grids $T_{i}=\sum_{u=0}^{i-1}\delta_{u}$
with $T_{0}=0$ and accrual factors $\{\delta_{u}|u=0,...,e-1\}$.
Additionally, $A$, and $S$ represent annuity factor and swap rate,
respectively. 

A Bermudan swaption, denoted as $B^{i}$, can be exercised at discrete
times $\{T_{u}|u=i,...,e-1\}$. In this study, we focus on a trade
in which, if exercised at $T_{u}$, the holder enters into a swap
$U^{u,e,K}$.\footnote{For simplicity, we assume that the exercise time and the swap start
time are the same. Extending this to cases where the exercise time
and swap start time differ is expected to be straightforward.}Thus, the underlying swap for the Bermudan swaption is assumed to
have a common maturity of $T_{e}$, and such swaps that share the
same maturity are referred to as coterminal swaps.

For hedging $B^{i}$, we consider coterminal swaps and swaptions as
the hedge assets. The coterminal swaps are represented by $\{U^{u,e,K}|u=i,...,e-1\}$.
A swaption $O^{u,j,K}$ is a European option with an underlying swap
$U^{u,j,K}$, where the exercise date aligns with the start date of
the underlying swap. Specifically, the set of swaptions with coterminal
underlying swaps, $\{O^{u,e,K}|u=i,...,e-1\}$, is referred to as
coterminal swaptions. Together, these coterminal swaps and swaptions
provide a basis for hedging the Bermudan swaption.

Henceforth, all assets are considered in terms of their relative prices
with respect to the discount factor at maturity $P_{t}^{e}$. Specifically,
we define the relative price of an asset $X_{t}$ \LyXZeroWidthSpace{}
as $\tilde{X}_{t}=\frac{X_{t}}{P_{t}^{e}}$, and, to simplify notation,
we denote this relative price as $X_{t}$. In this paper, we construct
arbitrage-free scenarios using a pricing measure that takes $P_{t}^{e}$
as the numéraire. This choice simplifies the approach discussed in
the paper, though other numéraires can be used within this framework.
For hedging strategy P\&L, this approach effectively holds the P\&L
generated in each period in a discount bond $P_{t}^{e}$.

\subsection{Overview of Scenario Generation Process}

As mentioned earlier, this paper adopts the Swap Market Bergomi Model
(SMBM) for arbitrage-free scenario generation. The process involves
the following steps:
\begin{enumerate}
\item Determine the exogenous model parameters of the SMBM: This step involves
defining parameters such as the initial forward variance swap rate,
correlation coefficients among state variable processes, volatility-of-volatility
factors, and mean-reversion parameters for the variance swap rate
process. These parameters can be estimated from the historical dynamics
of hedge assets or set according to the market outlook of the position
manager. A detailed discussion of this parameter estimation is outside
the scope of this paper. Once this step is completed, it does not
need to be repeated unless there is a significant change in the assumptions
used for estimation of the parameters.
\item Calculate the current state variables of the SMBM: The state variables
are derived from the current market prices of the hedge assets so
that the arbitrage-free model prices align with market prices. To
ensure accurate and feasible calculation, the number of state variables
is chosen to match the number of hedge assets. This step and all subsequent
steps will be performed daily.
\item Simulate future paths: Using the parameters and state variables obtained
in steps 1 and 2, we simulate future paths of the state variables.
This step can be executed using established Monte Carlo simulation
methods.
\item Calculate the prices of hedge assets: For each simulated path, we
compute the prices of hedge assets using the future values of the
state variables. The price of an interest rate swap can be calculated
analytically using the state variables. For swaptions, calculating
the conditional expectation is required, though pre-training techniques
can accelerate this computation for coterminal swaption valuation.
\end{enumerate}
In the next section, we discuss the dynamics of the SMBM and the price
calculation process for the hedge assets.

\subsection{Swap Market Bergomi Model}

The Swap Market Bergomi Model \cite{SMBM} is a term structure model
for interest rates that uses both the swap rate and its variance swap
rate as state variables. Swap market models typically offer flexibility
in choosing which set of swap rates to model, such as coterminal or
costart swap rates. In this paper, we define the joint dynamics of
coterminal swap rates and their associated variance swap rates.

The SMBM approach begins by modeling two-factor process for each swap
rate $S^{i,e}$, comprising the swap rate itself and its variance
swap rate under its own annuity measure $\mathcal{A}^{i,e}$. A change
of measure operation is then applied to construct the joint distribution
of the coterminal swap rates and variance swap rates. Let $\mathcal{M}^{i}$
represent the two-factor model for the swap rate $S^{i,e}$, and let
$\mathcal{M}^{Joint}$ denote the model for the joint distribution.
This setup allows efficient calculation of a coterminal swaption price
on $S^{i,e}$ from given state variables, as the computation depends
only on the dynamics of $\mathcal{M}^{i}$.

Moreover, for valuing coterminal swaptions, if pre-training is applied
using a regression model, the problem can be approached as a regression
task involving three parameters: the swap rate, the state variable
for variance swap rate, and the time to maturity, assuming the other
model parameters of $\mathcal{M}^{i}$ are fixed. Given this relatively
low dimensionality, high-precision regression methods, such as tensor
splines, can be effectively applied for accurate valuation.

\subsubsection*{Joint Dynamics of State Variables of the SMBM}

Following \cite{SMBM}, we outline the method used to construct the
dynamics. For detailed definitions and properties, please refer to
\cite{SMBM}. The SMBM construction begins with the formulation of
the stochastic differential equation (SDE) for $\mathcal{M}^{i}$.
Under the annuity measure $\mathcal{A}^{i,e}$, which is linked to
the annuity factor $A^{i,e}$, the swap rate process $S_{t}^{i,e}$
and the infinitesimal variance swap rate $\xi_{t}^{i,e,T}$ follow
these dynamics:
\begin{eqnarray}
dS_{t}^{i,e} & = & \sqrt{\xi_{t}^{i,e,t}}\,dW_{t}^{(i,e),\mathcal{A}^{i,e}}\\
dX_{t}^{i,e} & = & -\kappa^{i,e}X_{t}^{i,e}dt+dZ_{t}^{(i,e),\mathcal{A}^{i,e}}\\
\xi_{t}^{i,e,T} & = & \xi_{0}^{i,e,T}\exp\left[\omega^{i,e}e^{-\kappa^{i,e}(T-t)}X_{t}^{i,e}-\frac{1}{2}\left(\omega^{i,e}\right)^{2}e^{-2\kappa^{i,e}(T-t)}E_{0}^{\mathcal{A}^{i,e}}[\left(X_{t}^{i,e}\right)^{2}]\right]
\end{eqnarray}
Here $X_{t}^{i,e}$ is the state variable process for the variance
swap rate. $\omega^{i,e}$ and $\kappa^{i,e}$ are static model parameters,
representing the volatility-of-volatility factor and the mean-reversion
parameter for the variance swap rate process. $W_{t}^{(i,e),\mathcal{A}^{i,e}}$and
$Z_{t}^{(i,e),\mathcal{A}^{i,e}}$ represent correlated Brownian motions
under the annuity measure $\mathcal{A}^{i,e}$. 

After constructing $\mathcal{M}^{i}$ for each $i=1,...,e-1$, the
joint dynamics $\mathcal{M}^{Joint}$ is constructed through the following
steps:
\begin{enumerate}
\item Specify the pricing measure for $\mathcal{M}^{Joint}$: This paper
uses the terminal measure $\mathcal{P}^{e}$, which is associated
with the discount factor $P_{t}^{e}$. 
\item Specify the local correlation coefficients: Define the local correlation
coefficients for the following combinations: (a) between $dW_{t}^{(i,e),\mathcal{A}^{i,e}}$
for different $i$, (b) between $dZ_{t}^{(j,e),\mathcal{A}^{j,e}}$
for different $j$, and (c) between $dW_{t}^{(i,e),\mathcal{A}^{i,e}}$
and $dZ_{t}^{(j,e),\mathcal{A}^{j,e}}$ for all combinations of $i$
and $j$, except for $i=j$, which is defined during the construction
of $\mathcal{M}^{i}$. 
\item Replace the Brownian motions: Replace $W_{t}^{(i,e),\mathcal{A}^{i,e}}$
and $Z_{t}^{(i,e),\mathcal{A}^{i,e}}$ with the Brownian motions under
the measure $\mathcal{P}^{e}$, denoted as $W_{t}^{(i,e),\mathcal{P}^{e}}$
and $Z_{t}^{(i,e),\mathcal{P}^{e}}$, by calculating the arbitrage-free
drift. Details on the drift term expressions can be found in \cite{SMBM}.
\end{enumerate}

\subsubsection*{Mapping from State Variables of the SMBM to Hedge Asset Prices}

The arbitrage-free value of coterminal interest rate swap is given
by $A^{i,e}(K-S^{i,e})$. Since both $A^{i,e}$ and $S^{i,e}$ are
defined as functions of the discount factors $\{P^{u}|u=i,...,e-1\}$,
it is possible to calculate the sets of $\{P^{u}|u=i,...,e-1\}$ and
$\{A^{u,e}|u=i,...,e-1\}$ from the given set of $\{S^{u,e}|u=i,...,e-1\}$
as state variables, determining the value of the interest rate swap
analytically\footnote{Note that all asset prices are considered in terms of relative prices
to $P^{e}$.}. 

A swaption $O^{i,e,K}$, modeled by $\mathcal{M}^{i}$, is a European
option with a payoff of $A^{i,e}(K-S^{i,e})^{+}$ at maturity $T_{i}$.
Its arbitrage-free value is given by $A_{t}^{i,e}E_{t}^{\mathcal{A}^{i,e}}[(K-S_{T_{i}}^{i,e})^{+}]$
for $t\leq T_{i}$. Using the dynamics of the $\mathcal{M}^{i}$,
this expectation value can be calculated\footnote{For an efficient computational method using one-dimensional Monte
Carlo simulations, see \cite{SMBM}.}. Thus, the price of the swaption at any time $t\leq T_{i}$ can be
obtained as a function of three variables: $S_{t}^{i,e}$, $X_{t}^{i,e}$,
and $t$.

Once a regression model of the swaption price $O^{i,e,K}$ with respect
to $S_{t}^{i,e}$, $X_{t}^{i,e}$, and $t$ is constructed, mapping
from state variables to the swaption price $O^{i,e,K}$ becomes very
efficient in the market scenario path generation. Given the low dimensionality
of the input (only three variables), a high-accuracy regression model,
such as tensor splines, can be used.

\section{Deep Hedging Bermudan Swaptions}

\subsection{Deep Hedging Approach}

In the deep hedging approach \cite{DeepHedgeOriginal}, the optimal
hedging and pricing problem is formulated as a minimization problem,
where the objective is to minimize a given convex risk measure function
$\rho$ applied to the distribution of the hedged P\&L. More specifically,
the optimal hedging strategy $w^{*}$ is defined as:

\begin{equation}
w^{*}=\text{argmin}{}_{w\in\mathcal{H}}\rho(\text{PL}(w,Z))\label{eq:DHOptimization}
\end{equation}

Here, $\mathcal{H}$ denotes the set of possible hedging strategies,
and $\text{PL}$ represents the distribution of hedged P\&L, which
is calculated as follows\footnote{In this paper, we consider a case where a trader holds the Bermudan
swaption instead of selling it. Therefore, the discussion here reverses
the sign of the payoff compared to the original paper \cite{DeepHedgeOriginal}.}:
\begin{equation}
\text{PL}(w,Z)=Z+\sum_{k=0}^{n-1}w_{k}\cdot(I_{k+1}-I_{k})-\sum_{k=0}^{n}c_{k}(w_{k}-w_{k-1})\label{eq:DHFormulation}
\end{equation}

In this equation, $Z$ represents the payoff value distribution of
the trade being hedged, while $w$, $I$ and $c$ denote the hedge
quantity vector, the price vector of the hedge assets, and the hedging
cost function, respectively. Trading days span from $t_{0}=0$\LyXZeroWidthSpace{}
to $t_{n}$, where $t_{n}$ \LyXZeroWidthSpace is the maturity date
of the trade. It is also assumed that $w_{-1}=w_{n}=0$.

In this approach, the price $V$ of the derivative payoff $Z$ is
determined as the indifference price. Specifically, if we denote $\rho(\text{PL}(w,Z))$
with the optimal hedging strategy as $\pi(Z)=\min_{w\in\mathcal{H}}\rho(\text{PL}(w,Z))$,
the price $V$ is calculated as the value satisfying $\pi(Z-V)=\pi(0)$.
This represents the fair cash amount that the hedger considers appropriate
to pay to enter the position, accounting for risk. By using the cash-invariance
property of the convex risk measure function, this condition can be
reformulated as:

\begin{equation}
V=-\pi(Z)+\pi(0)\label{eq:ModelPriceDefinition}
\end{equation}

Here, $\pi(0)$ is a term associated with the value of statistical
arbitrage strategy. Since this study uses an arbitrage-free scenario
generator, this value is theoretically expected to be zero. Therefore,
in numerical calculations below, we proceed with the assumption $\pi(0)=0$.

In the deep hedging approach, the optimization problem in (\ref{eq:DHOptimization})
is solved by approximating $w$ with an artificial neural network
(ANN) and using a neural network optimization framework.

\subsection{Deep Hedging Approach for Bermudan Swaptions by Component Decomposition}

\subsubsection{Construction of Deep Hedging Model Using Decomposition Formula}

The Bermudan swaption considered in this paper grants the holder the
right to enter into a coterminal interest rate swap $U^{i,e,K}$ \LyXZeroWidthSpace{}
at the exercise time $T_{i}$. Replacing the underlying asset $U^{i,e,K}$
at the $i$-th exercise of the Bermudan swaption\LyXZeroWidthSpace{}
with $O^{i,e,K}$ does not change its economic properties. While this
replacement is mathematically straightforward, rewriting the payoff
of the Bermudan swaption in this form allows us to explicitly represent
it as a type of exchange option between swaptions.

To construct the deep hedging model for this Bermudan swaption, we
follow the known decomposition method \cite{AndersenPiterbarg}. We
first decompose the Bermudan swaption $B^{1}$ into components $\{C^{i}|i=1,...,e-1\}$
with the following definitions:
\begin{enumerate}
\item The right to receive $O^{e-1,e,K}$ at $T_{e-1}$ is defined as $C^{e-1}.$
\item The right to receive $O^{i,e,K}$ and pay $B^{i+1}$ (in other words,
the right to receive option spread $O^{i,e,K}-B^{i+1})$ at $T_{i}$
is defined as $C^{i}$ for $i\in\{1,...,e-2\}$.
\end{enumerate}
Then, we can express $B^{i}$ as the sum of $C^{i}$ from $i$ to
$e-1$, that is, $B^{i}=\sum_{u=i}^{e-1}C^{u}$.

In this study, we construct the hedging model for $\ensuremath{B^{i}}$
in a reverse order from $\ensuremath{i=e-1}$ to $\ensuremath{i=1}$.
Assuming that the hedging model for $\ensuremath{B^{i+1}}$ has already
been constructed, we then use this model as a component of the hedging
strategy model for $\ensuremath{B^{i}}$. Since $\ensuremath{B^{i}=B^{i+1}}+C^{i}$,
constructing the hedging model for $\ensuremath{B^{i}}$, given the
hedging model for $B^{i+1}$, effectively involves constructing the
hedging model for $C^{i}$.

$\ensuremath{C^{i}}$ is a European option that grants the right to
receive $O^{i,e,K}$ and pay $\ensuremath{B^{i+1}}$ at maturity $\ensuremath{T^{i}}$.
To determine the value of $\ensuremath{B^{i+1}}$ at $T_{i}$, we
need to use the deep hedging pricing framework again. However, a straightforward
approach would require market simulations based on the every state
variables at $T_{i}$, which is impractical. To address this, we use
an ANN to model the exercise decision.

\subsubsection{Hedging Models and Early Exercise Decision Models}

Building on the previous discussion, we now construct a concrete deep
hedging method for the Bermudan swaption, introducing both the hedging
model and the early exercise decision model. We assume that the hedging
model for $\ensuremath{B^{i}}$ for $\ensuremath{t<T_{i}}$ is defined
as:
\begin{equation}
\mathcal{N}^{BH,i}=\mathcal{N}^{CH,i}+\mathcal{N}^{BH,i+1}
\end{equation}

Also, the hedging model for $t\geq T_{i}$\LyXZeroWidthSpace{} is
set as $\mathcal{N}^{BH,i}=\mathcal{N}^{BH,i+1}$, which is used if
$\ensuremath{B^{i}}$ is not exercised at $t=T_{i}$. Additionally,
we assume $\mathcal{N}^{BH,e}=0$. Furthermore, at $t=T_{i}$, we
define the early exercise decision model as $\ensuremath{\mathcal{N}^{E,i}}$.
Here, $\mathcal{N}^{BH,i}$, $\mathcal{N}^{CH,i}$, and $\mathcal{N}^{E,i}$
represent the ANN models for the hedging strategy for $B^{i}$, $C^{i}$,
and early exercise decision, respectively.

$\mathcal{N}^{CH,i}$ outputs the hedge amounts for the hedge assets,
which will be described in Section 3.2.5. $\mathcal{N}^{E,i}$ outputs
the probability of exercising at time $T_{i}$. A sigmoid activation
function is applied to the output layer of $\mathcal{N}^{E,i}$ to
ensure values fall between 0 and 1.

The input features for the hedging model $\mathcal{N}^{CH,i}$ include
the time $t$ and the prices of the coterminal swaps, $\{U_{t}^{u,e,K}|u=i,...,e-1\}$.
Additionally, for strategies other than $\mathcal{S}^{I}$ (described
below), the input features include the prices of the coterminal swaptions,
$\{O_{t}^{u,e,K}|u=i,...,e-1\}$. For the exercise decision model
$\mathcal{N}^{E,i}$, the input features consist of the prices of
the coterminal swaps at $T_{i}$, $\{U_{T_{i}}^{u,e,K}|u=i,...,e-1\}$,
and, for strategies other than $\mathcal{S}^{I}$, the coterminal
swaptions,$\{O_{T_{i}}^{u,e,K}|u=i,...,e-1\}$.

As described earlier, during the training of the hedging model for
$\ensuremath{B^{i}}$, $\mathcal{N}^{BH,i+1}$ is assumed to be available.
Therefore, during the training process for $\ensuremath{B^{i}}$,
the models $\mathcal{N}^{CH,i}$ and $\mathcal{N}^{E,i}$ are optimized.

\subsubsection{Training Payoff for $\ensuremath{B^{i}}$}

In the context of hedging for $\ensuremath{B^{i}}$, the P\&L components
controlled by $\mathcal{N}^{CH,i}$ and $\mathcal{N}^{E,i}$ are relevant
only up to $T_{i}$. Therefore, $B^{i}$ can be trained as if it were
a European option with a payoff at $T_{i}$, which is calculated as
the sum of the following two terms:
\begin{itemize}
\item $\mathcal{N}^{E,i}\cdot O^{i,e,K}$ : This term represents the expected
exercise value, weighted by the exercise probability.
\item $(1-\mathcal{N}^{E,i})\cdot V^{Cont,i}$: This term represents the
pathwise continuation value, weighted by the non-exercise probability.
\end{itemize}
Here, $V^{Cont,i}$ refers to the hedged P\&L defined in (\ref{eq:DHFormulation})
generated for $t\geq T_{i}$ \LyXZeroWidthSpace by using each simulated
path of the state variables and the hedging strategy based on $\mathcal{N}^{BH,i+1}$,
with $V^{Cont,e-1}=0$. Specifically, $V^{Cont,i}$ satisfies the
following recursive equation:

\begin{equation}
V^{Cont,i}=\mathcal{N}^{E,i+1}\cdot O^{i+1,e,K}+(1-\mathcal{N}^{E,i+1})\cdot V^{Cont,i+1}+\sum_{k}w_{k}^{*}\cdot(I_{k+1}-I_{k})
\end{equation}

In the equation, $w_{k}^{*}$ represents the optimized hedge amounts
obtained from $\mathcal{N}^{BH,i+1}$, $\{I_{k}\}$ denotes the price
vector of the hedge assets, and the summation over $k$ spans the
time grid $t\in[T_{i},T_{i+1})$. Note that $V^{Cont,i}$ is calculated
using information from only that single path.

The payoff can also be rewritten as $\mathcal{N}^{E,i}(O^{i,e,K}-V^{Cont,i})+V^{Cont,i}$.
Here, the first term corresponds to the payoff $(O^{i,e,K}-B^{i+1})^{+}$of
$\ensuremath{C^{i}}$, and the second term corresponds to $B^{i+1}$,
where $B^{i+1}$ is approximated by the single-path calculation $V^{Cont,i}$.

\subsubsection{Hedging Cost}

In this paper, we simplify the analysis by excluding hedging costs.
However, in principle, incorporating hedging costs into this framework
is feasible. Note that, while each individual hedging model is trained
separately, it is essential to properly account for the netting effect
of hedging costs across the components\textquoteright{} individual
models during the training process.

\subsubsection{Hedging Strategies}

This section describes the hedging strategies analyzed in this study.
In this paper, a \textquotedbl{}hedging strategy\textquotedbl{} refers
to the selection of hedge assets and the constraints on hedge amounts.
The primary strategies proposed in this paper are the Option Spread
Strategy $\mathcal{S}^{OS}$ and the Trade Strike Interest Rate Swap
and Swaption Hedge Strategy $\mathcal{S}^{I+S}$. For $\mathcal{S}^{OS}$,
we also explore an interesting subset strategy, $\mathcal{S}^{Max}$.
Additionally, we include $\mathcal{S}^{I+S,M}$ and $\mathcal{S}^{I}$
as reference strategies in the numerical analysis.

\paragraph*{Option Spread Strategy $\mathcal{S}^{OS}$}

In this strategy, the hedge asset for $\mathcal{N}^{CH,i}$ is a single
asset, $G^{i}=O^{i,e,K}-H^{i+1}$. Here, $H^{i}=\sum_{u=i}^{e-1}w_{u}^{*}G^{u}$
represents the optimal hedging portfolio for \textbf{$B^{i}$}, where
$H^{e}=0$ and $w_{u}^{*}$ is the optimal hedge amount to short determined
through the training of $\mathcal{N}^{CH,u}$.

According to this definition, $H^{i}=\sum_{u=i}^{e-1}q_{u}^{i}O^{u,e,K}$,
where $q_{u}^{i}=w_{u}^{*}\prod_{k=i}^{u-1}(1-w_{k}^{*})$\footnote{The value of an empty product is defined as 1.}.
A key observation here is that $\sum_{u=i}^{e-1}q_{u}=1$, meaning
that $H^{i}$ is a portfolio of swaptions with a total weight of 1\footnote{Note that $w_{e-1}^{*}=1$.}.
Furthermore, by applying the constraint $0\leq w_{u}^{*}\leq1$, $0\leq q_{u}\leq1$
consequently holds. This constraint is enforced only after training,
allowing unrestricted values for the hedge amounts during training.
However, after training is complete, the amounts are clipped to ensure
they fall within the range of 0 to 1.

\subparagraph*{Economic Implications of the $\mathcal{S}^{OS}$ Hedging Strategy}

Constructing the hedging model for $B^{i}$, given the hedging model
for $B^{i+1}$, essentially involves creating the hedging model for
$\ensuremath{C^{i}}$, which is a European option that expires at
$T_{i}$ with a payoff of $(O^{i,e,K}-B^{i+1})^{+}$. Letting $X^{i}=O^{i,e,K}-B^{i+1}$,
the resulting payoff $\left(X^{i}\right)^{+}$ represents a European
option on the single asset $X^{i}$. Thus, we can interpret $\mathcal{S}^{OS}$
as a form of \textquotedbl{}delta hedging\textquotedbl{} a European
option on $X^{i}$ by using $X^{i}$ itself. Since $B^{i+1}$ is not
directly tradable, we substitute its hedging portfolio $H^{i+1}$
for $B^{i+1}$ in $X^{i}$, resulting in $G^{i}=O^{i,e,K}-H^{i+1}$
as the hedge asset.

\subparagraph*{Economic Meaning of \textquotedbl{}Delta Hedging\textquotedbl{} a
European Option on $X^{i}$}

The option being hedged in this case is essentially an option on an
option spread, or an \textquotedblleft option on an option.\textquotedblright{}
This type of option has two different types of optionality: (1) the
optionality of the underlying asset (an option), and (2) the optionality
of the option on the underlying option. The $\mathcal{S}^{OS}$ strategy
hedges only the first type of optionality\textemdash the optionality
of the underlying option\textemdash while leaving the second type
unhedged.

In Bermudan swaption model analysis, this second type of optionality
is often referred to as a \textquotedblleft switch option\textquotedblright{}
or \textquotedblleft Bermudanality\textquotedblright \cite{AndersenPiterbarg}.
Using this terminology, the $\mathcal{S}^{OS}$ strategy can thus
be described as hedging only the underlying swaption\textquoteright s
optionality, leaving the switch option unhedged.

\subparagraph*{Rationale for Not Hedging the Switch Option}

Theoretically, avoiding hedging the switch option may appear sub-optimal
since $\mathcal{S}^{OS}$ is a subset of more flexible strategies
like $\mathcal{S}^{I+S}$, discussed below. However, this choice is
motivated by the goal of limiting downside risk. As previously mentioned,
the $\mathcal{S}^{OS}$ strategy resembles delta hedging a European
option. A long position in delta-hedged European options typically
results in a gamma-long position, thereby limiting downside risk.

\subparagraph*{Intuitive Economic Interpretation}

Another advantage of the $\mathcal{S}^{OS}$ strategy is that the
hedge weights produced by the trained model offer an intuitive economic
interpretation. The hedging portfolio is a combination of European
swaptions with positive weights that sum to 1. This allows us to view
the Bermudan swaption as a weighted combination of swaptions, where
the weights reflect an \textquotedblleft exercise probability\textquotedblright{}
under a specific probability measure. This interpretation provides
a useful means of intuitively validating hedging model outputs.

\paragraph*{Option Spread Max Strategy $\mathcal{S}^{Max}$}

We would also like to introduce the $\mathcal{S}^{Max}$ strategy
as an interesting subset of the $\mathcal{S}^{OS}$ strategy. In other
words, the $\mathcal{S}^{OS}$ strategy can be seen as an extension
of this simple approach, enhanced by deep hedging techniques.

In the $\mathcal{S}^{Max}$ strategy, the hedge amount to short $G^{i}$
is restricted to either 0 or 1, depending solely on the sign of $G^{i}$.
Specifically, $w_{i}^{*}=\Theta(G^{i})$, where $\Theta$ is the Heaviside
function. This hedging model can be implemented analytically and no
training processes.

The Bermudan hedge portfolio to short is given as $H^{i}=\sum_{u=i}^{e-1}q_{u}^{i}O^{u,e,K}$,
where $q_{u}^{i}=1$ for $u=\arg\max_{j\in\{i,...,e-1\}}O^{j,e,K}$
and otherwise 0. Essentially, this strategy shorts the swaption with
the highest value at each step. It serves as a simple yet effective
benchmark hedging model for comparison with more sophisticated strategies.

In this approach, we also adopt a straightforward exercise strategy:
the option is exercised at time $T_{i}$ \LyXZeroWidthSpace only if
$O^{i,e,K}$ has the highest option value in $\{O^{u,e,K}|u=i,...,e-1\}$.
This exercise strategy can also be implemented analytically.

\paragraph*{Trade Strike Interest Rate Swap and Swaption Hedge Strategy $\mathcal{S}^{I+S}$}

In this strategy, the hedge assets for $\mathcal{N}^{CH,i}$ consist
of a set of single-strike coterminal interest rate swaps $\{U^{u,e,K}|u=i,...,e-1\}$
\LyXZeroWidthSpace{} and swaptions $\{O^{u,e,K}|u=i,...,e-1\}$, allowing
unrestricted hedge amounts. Economically, this strategy hedges the
optionalities of both the underlying and the switch options.

\paragraph*{Multiple Strikes Interest Rate Swap and Swaption Hedge Strategy $\mathcal{S}^{I+S,M}$}

In this strategy, the hedge assets for $\mathcal{N}^{CH,i}$ consist
of a set of single-strike coterminal interest rate swaps $\{U^{u,e,K}|u=i,...,e-1\}$
\LyXZeroWidthSpace{} and multiple-strike coterminal swaptions $\{O^{u,e,k}|u=i,...,e-1,k=K_{1},K_{2},...\}$,
allowing unrestricted hedge amounts. While it is possible to compute
the hedge assets on the generated state space, the state space does
not fully capture the dynamics of multiple-strike swaptions, specifically
as it lacks the dynamics of swaption skew. Therefore, this strategy
is considered only as a reference in this study, aimed at understanding
the change in hedge performance when increasing the number of strikes
for hedge swaptions.

\paragraph*{Interest Rate Swap Only Hedge Strategy $\mathcal{S}^{I}$}

In this strategy, the hedge assets for $\mathcal{N}^{CH,i}$ consist
of a set of single-strike coterminal interest rate swaps $\{U^{u,e,K}|u=i,...,e-1\}$,
allowing unrestricted hedge amounts. This strategy corresponds to
situations where swaptions are not available for trading in the market
as hedge assets. It serves as a reference to assess the effectiveness
of swaption hedging.

\subsubsection{Training and Daily Operation Using the Trained Model}

During training, both the hedging model $\mathcal{N}^{CH,i}$ and
the exercise decision model $\ensuremath{\mathcal{N}^{E,i}}$ are
optimized simultaneously. To prevent over-fitting $\mathcal{N}^{CH,i}$
and $\ensuremath{\mathcal{N}^{E,i}}$ to the training paths, several
common machine learning regularization techniques are applied, including
dropout, stochastic batch sampling, and splitting data into distinct
training and evaluation sets.

Once the model has been trained, as long as the state variables remain
within the training set's range and there are no significant changes
to the model assumptions of the SMBM used for generating the training
set, daily retraining of the model is not necessary. Instead, we proceed
from Step 2 in Section 2.2, generating scenarios based on the current
market prices of the hedge assets. This allows us to calculate hedge
quantities, revise estimates of the hedged P\&L distribution, and
update the model indifferece price.

\section{Numerical Experiment}

\subsection{Calculation Assumptions}

\subsubsection{The Trade Under Consideration}

The trade under consideration is a Bermudan swaption with a fixed
strike rate of 2\%. There are five exercise opportunities, with the
first exercise occurring four years after the initial trade date $t=0$,
followed by annual exercise opportunities. Specifically, the exercise
dates are at $T_{i}=i+3$ for $i=1,2,3,4,5$. The underlying swap
has a common maturity of $T_{6}=T_{e}=9$, and both the funding leg
and the coupon leg of the swaps have an annual coupon frequency.

\subsubsection{Model Parameters for the SMBM for Scenario Generation}

The parameters for the SMBM scenario generation model are as follows.
$S_{0}^{i,e,K}$, $X_{0}^{i,e}$, $\kappa^{i,e}$, and $\omega^{i,e}$
are constant for $i=1,2,3,4,5$ with $S_{0}^{i,e,K}$=0.02, $X_{0}^{i,e}$=0,
$\kappa^{i,e}=0$, and $\omega^{i,e}=0.8403$. $\xi_{0}^{i,e,T}$
\LyXZeroWidthSpace is assumed to be independent of $T$, meaning $\xi_{0}^{i,e,T}=\xi_{0}^{i,e}$,
with $\xi_{0}^{i,e}$ values as:
\begin{equation}
\{\xi_{0}^{i,e}\}_{i=1,2,3,4,5}=[1.132e-04,1.228e-04,1.287e-04,1.344e-04,1.451e-04]
\end{equation}

For correlation, distinct correlation matrices were applied for each
diffusion period, with the specific values provided in the Appendix.

\subsubsection{Scenario Generation Simulation Parameters}

The time grid interval is set to 1/32, and both the training and test
sets consist of 4096 paths.

\subsubsection{Deep Hedging Training and Evaluation Setup}

The neural network structure is shared between the hedging and exercise
decision models, with the output layer of the exercise decision model
using a sigmoid function. Each model is a fully connected network
with four hidden layers, each containing 32 units. Batch normalization
and dropout (with a rate of 0.5) are applied to each hidden layer.
The Adam optimizer is used. The Conditional Value at Risk (CVaR) is
used as the convex risk measure, defined as:
\begin{eqnarray}
\text{CVaR}_{\alpha}(X) & = & \frac{1}{1-\alpha}\int_{0}^{1-\alpha}\text{VaR}_{\gamma}(X)d\gamma\\
\text{VaR}_{\gamma}(X) & = & \inf\{m\in\mathbb{R}:P(X<-m)\leq\gamma\}
\end{eqnarray}
 The confidence level $\alpha$ is set to 0.2, 0.4, 0.6, and 0.8,
and the hedging strategies used include the five strategies discussed
in Chapter 3: $\mathcal{S}^{OS}$, $\mathcal{S}^{Max}$, $\mathcal{S}^{I+S}$,
$\mathcal{S}^{I}$, and $\mathcal{S}^{I+S,M}$. For $\mathcal{S}^{I+S,M}$,
hedge swaption strikes of 0.01, 0.015, and 0.02 are used, while in
all other cases, the hedge asset strikes are set to 0.02, the same
as the Bermudan swaption strike. This setup yields 20 distinct models
for training and evaluation.

\subsection{Results}

The values in this section are expressed as a ratio to the notional
amount and shown in basis points.

\subsubsection{Hedged P\&L Distribution}

Table \ref{tab:pv} presents the evaluation metrics and related indicators
for the valuation results of the 20 different settings at $t=0$,
based on the test set paths. Key columns include:
\begin{itemize}
\item \textbf{Model Value}: The indifferent price calculated using (\ref{eq:ModelPriceDefinition}).
\item \textbf{NonArb Value}: The expected value of the unhedged P\&L, corresponding
to the arbitrage-free price when the trained exercise decision model
is used.
\item \textbf{Hedge PnL Mean}: The average P\&L from the hedge position. 
\item \textbf{Model Switch Value} and \textbf{NonArb Switch Value}: Calculated
by subtracting the maximum swaption value at $t=0$ from the \textbf{Model
Value} and \textbf{NonArb Value}, representing the model value metric
of the switch option in the Bermudan swaption. Here, the maximum swaption
value at $t=0$ is $O_{0}^{u,e,K}$\LyXZeroWidthSpace{} with $u=\arg\max_{j\in\{1,2,3,4,5\}}O_{0}^{j,e,K}$.
\end{itemize}
Table \ref{tab:dist} provides details on the distribution of P\&L
drawdowns, defined as the differences between \textbf{Model Value}
and the hedged P\&L. Key columns include:
\begin{itemize}
\item \textbf{P25}, \textbf{P50}, and \textbf{P75:} The 25th, 50th, and
75th percentiles of the P\&L drawdown distribution.
\item \textbf{IQR:} The interquartile range, calculated as the difference
between P75 and P25.
\item \textbf{Loss Prob}: The probability that the P\&L drawdown is negative
\item \textbf{Expected Loss}: The average loss when a loss occurs.
\item \textbf{CVaR95} and \textbf{CVaR99}: CVaR at $\alpha=0.95$ and $\alpha=0.99$,
providing measures of downside risk.
\end{itemize}
In the following, \textquotedbl{}the swaption hedging strategies\textquotedbl{}
refers to $\mathcal{S}^{OS}$, $\mathcal{S}^{Max}$, $\mathcal{S}^{I+S}$,
and $\mathcal{S}^{I+S,M}$.

\paragraph{Key Insights from the Results}

\subparagraph{Impact of Confidence Level $\alpha$ on Evaluation:}

Across all strategies, increasing the confidence level $\alpha$ leads
to more conservative evaluations. For the swaption hedging strategies,
as $\alpha$ approaches 1, the Switch Value decreases toward zero.
Conversely, as $\alpha$ decreases, the evaluation approaches a arbitrage-free
valuation. Thus, in this framework, $\alpha$ acts as a parameter
controlling the degree of recognition of the switch value of the arbitrage-free
pricing.

\subparagraph{Reduction in Downside Risk:}

A higher confidence level $\alpha$ tends to reduce downside risks,
as shown by CVaR95 and CVaR99, particularly for the $\mathcal{S}^{I+S}$
and $\mathcal{S}^{I+S,M}$ strategies. For $\mathcal{S}^{OS}$ and
$\mathcal{S}^{Max}$, the increase in downside risk with a lower $\alpha$
is less pronounced. This demonstrates that $\alpha$ can be used to
manage downside risk, and that $\mathcal{S}^{OS}$ and $\mathcal{S}^{Max}$
strategies are more effective in reducing downside risk across a wide
range of $\alpha$ values.

\subparagraph{Reduction in P\&L Volatility:}

The ability of each strategy to reduce P\&L volatility through hedging
can be assessed using the IQR values. Strategies $\mathcal{S}^{I+S}$
and $\mathcal{S}^{I+S,M}$ generally perform better than $\mathcal{S}^{OS}$
and $\mathcal{S}^{Max}$ as expected, because $\mathcal{S}^{OS}$
and $\mathcal{S}^{Max}$ are subset strategies of $\mathcal{S}^{I+S}$
and $\mathcal{S}^{I+S,M}$. Meanwhile, the IQR for $\mathcal{S}^{OS}$
and $\mathcal{S}^{Max}$ significantly improves upon $\mathcal{S}^{I}$,
confirming that the hedging performance for the Bermudan swaption
improves substantially with the introduction of swaptions as hedge
assets, even in strategies with restrictions on hedge amounts, like
$\mathcal{S}^{OS}$ and $\mathcal{S}^{Max}$.

\subparagraph*{Martingale Condition for Hedge Assets:}

The Hedge PnL Mean represents the expected P\&L obtained from the
hedging position and is generally a slightly positive across the swaption
hedging strategies. Scenarios generated by an arbitrage-free scenario
generator are used for training and evaluation alike. However, due
to numerical errors in SDE discretization and the limited number of
paths, ideal arbitrage-free conditions are not fully achieved in these
scenarios. As a result, some distributional bias remains in the evaluations
result, though its quantitative impact is minor.

\subparagraph{Framework's Value for Risk Management:}

This framework enables position managers to compare loss probabilities
and distributions across different model configurations, facilitating
more informed decision-making based on their risk tolerance.

\begin{table} \caption{Model Values and Model Risk Metrics (unit: basis points)} \label{tab:pv} \begin{tabular}{p{1.5cm}p{1.5cm}p{1.5cm}p{1.5cm}p{1.5cm}p{1.5cm}p{1.5cm}} \toprule  &  & Model Value & NonArb Value & Hedge PnL Mean & Model Switch Value & NonArb Switch Value \\ Strategy & $\alpha$ &  &  &  &  &  \\ \midrule \multirow{4}{*}{$\mathcal{S}^{I}$} & 0.2 & 345.2 & 456.2 & -1.7 & -51.8 & 59.1 \\  & 0.4 & 290.9 & 452.9 & -1.2 & -106.2 & 55.8 \\  & 0.6 & 245.0 & 452.3 & -3.1 & -152.0 & 55.2 \\  & 0.8 & 189.2 & 451.0 & -1.2 & -207.9 & 54.0 \\ \cline{1-7} \multirow{4}{*}{$\mathcal{S}^{Max}$} & 0.2 & 429.6 & 455.8 & 5.2 & 32.5 & 58.8 \\  & 0.4 & 415.9 & 455.8 & 5.2 & 18.8 & 58.8 \\  & 0.6 & 405.8 & 455.8 & 5.2 & 8.8 & 58.8 \\  & 0.8 & 397.8 & 455.8 & 5.2 & 0.7 & 58.8 \\ \cline{1-7} \multirow{4}{*}{$\mathcal{S}^{OS}$} & 0.2 & 434.1 & 456.3 & 4.5 & 37.1 & 59.3 \\  & 0.4 & 422.9 & 456.1 & 4.8 & 25.8 & 59.0 \\  & 0.6 & 414.7 & 455.6 & 5.4 & 17.6 & 58.6 \\  & 0.8 & 407.2 & 456.3 & 4.4 & 10.1 & 59.3 \\ \cline{1-7} \multirow{4}{*}{$\mathcal{S}^{I+S}$} & 0.2 & 444.1 & 456.5 & 6.3 & 47.0 & 59.5 \\  & 0.4 & 434.9 & 456.5 & 4.7 & 37.9 & 59.5 \\  & 0.6 & 420.9 & 456.5 & 3.1 & 23.9 & 59.5 \\  & 0.8 & 405.6 & 455.8 & 2.3 & 8.6 & 58.7 \\ \cline{1-7} \multirow{4}{*}{$\mathcal{S}^{I+S,M}$} & 0.2 & 447.2 & 456.7 & 8.7 & 50.1 & 59.6 \\  & 0.4 & 436.0 & 456.2 & 5.6 & 39.0 & 59.2 \\  & 0.6 & 420.1 & 456.0 & 2.5 & 23.1 & 59.0 \\  & 0.8 & 405.6 & 455.5 & 0.8 & 8.5 & 58.4 \\ \cline{1-7} \bottomrule \end{tabular} \end{table} 

\begin{table} \caption{Distribution Metrics for Initial Model Value and Realized PnL Differences (unit: basis points, excluding Loss Prob)} \label{tab:dist} \begin{tabular}{p{1.2cm}p{1.2cm}p{1.2cm}p{1.2cm}p{1.2cm}p{1.2cm}p{1.2cm}p{1.2cm}p{1.2cm}p{1.2cm}} \toprule  &  & P25 & P50 & P75 & IQR & Loss Prob & Expected Loss & CVaR95 & CVaR99 \\ Strategy & $\alpha$ &  &  &  &  &  &  &  &  \\ \midrule \multirow{4}{*}{$\mathcal{S}^{I}$} & 0.2 & -73.7 & 46.3 & 215.4 & 289.1 & 0.405 & 110.3 & 260.0 & 334.0 \\  & 0.4 & -16.0 & 92.4 & 254.8 & 270.8 & 0.284 & 77.3 & 172.3 & 243.2 \\  & 0.6 & 32.8 & 138.8 & 301.6 & 268.8 & 0.180 & 61.0 & 123.8 & 184.9 \\  & 0.8 & 86.4 & 192.0 & 358.8 & 272.3 & 0.083 & 44.3 & 65.2 & 123.4 \\ \cline{1-10} \multirow{4}{*}{$\mathcal{S}^{Max}$} & 0.2 & -20.8 & 5.9 & 52.2 & 73.0 & 0.449 & 21.5 & 32.5 & 32.5 \\  & 0.4 & -7.1 & 19.6 & 65.9 & 73.0 & 0.320 & 13.5 & 18.8 & 18.8 \\  & 0.6 & 2.9 & 29.6 & 75.9 & 73.0 & 0.219 & 7.4 & 8.8 & 8.8 \\  & 0.8 & 11.0 & 37.7 & 84.0 & 73.0 & 0.166 & 0.7 & 0.7 & 0.7 \\ \cline{1-10} \multirow{4}{*}{$\mathcal{S}^{OS}$} & 0.2 & -18.3 & 7.5 & 49.2 & 67.5 & 0.434 & 21.8 & 42.5 & 48.9 \\  & 0.4 & -4.8 & 19.2 & 55.2 & 60.0 & 0.302 & 14.2 & 28.8 & 34.7 \\  & 0.6 & 4.5 & 22.4 & 61.3 & 56.7 & 0.187 & 8.5 & 16.1 & 20.9 \\  & 0.8 & 11.1 & 28.5 & 68.2 & 57.2 & 0.089 & 5.3 & 8.0 & 12.2 \\ \cline{1-10} \multirow{4}{*}{$\mathcal{S}^{I+S}$} & 0.2 & -2.9 & 24.7 & 48.4 & 51.3 & 0.270 & 54.3 & 173.0 & 362.2 \\  & 0.4 & 4.4 & 21.7 & 43.4 & 39.1 & 0.208 & 28.6 & 77.9 & 155.2 \\  & 0.6 & 9.7 & 23.5 & 47.8 & 38.1 & 0.131 & 20.7 & 43.3 & 93.5 \\  & 0.8 & 9.7 & 23.0 & 55.7 & 46.1 & 0.074 & 6.9 & 9.8 & 21.0 \\ \cline{1-10} \multirow{4}{*}{$\mathcal{S}^{I+S,M}$} & 0.2 & -3.0 & 24.4 & 48.2 & 51.1 & 0.271 & 53.3 & 173.1 & 355.0 \\  & 0.4 & 4.7 & 21.9 & 41.0 & 36.4 & 0.201 & 29.8 & 81.3 & 168.3 \\  & 0.6 & 9.1 & 20.8 & 42.8 & 33.7 & 0.135 & 18.6 & 39.4 & 83.8 \\  & 0.8 & 8.8 & 21.0 & 54.2 & 45.5 & 0.067 & 8.4 & 10.8 & 25.9 \\ \cline{1-10} \bottomrule \end{tabular} \end{table}

\subsubsection{Cross-Parameter Validation of Hedging Model Resilience}

In this section, we assess the robustness of the hedging model trained
in the previous section by testing it on scenarios generated with
modified parameter sets of the SMBM. Specifically, we create 4096-path
reference scenarios where the signs of all rate-volatility correlations\textemdash namely,
those between $S^{i,e}$ and $X^{j,e}$\textemdash are reversed, and
the volatilty-of-volatlity (vol-vol) factors $\{\omega^{i,e}|i=1,2,3,4,5\}$
are reduced to half of their original values. The values of $\{X_{0}^{i,e}|i=1,2,3,4,5\}$\LyXZeroWidthSpace{}
are adjusted so that the coterminal swaption prices $\{O^{i,e,K}|i=1,2,3,4,5\}$
at $t=0$ align with the original scenario. All other state variables
and model parameters are kept at their previously used values. Subsequently,
using the model trained in the previous section, we conducted calculations
using the reference scenarios only for the evaluation step.

Table \ref{tab:pv_low_volvol} presents the evaluation metrics and
related indicators, while Table \ref{tab:dist_low_volvol} shows the
information on the distribution of P\&L drawdowns.

\paragraph{Key Insights from the Results}

\subparagraph{Valuation Result Comparison to the Original Setting:}

Overall, the swaption hedging strategies show little change in results,
indicating that the learned model demonstrates robustness. A closer
look reveals the following:
\begin{itemize}
\item Model Value: The Model Values in swaption hedging strategies remain
stable, particularly in cases with high $\alpha$ values. Specifically,
$\mathcal{S}^{OS}$ \LyXZeroWidthSpace{} and $\mathcal{S}^{Max}$
yield similar values across a wide range of $\alpha$, indicating
strong robustness.
\item IQR: $\mathcal{S}^{I+S}$ and $\mathcal{S}^{I+S,M}$ still generally
outperform $\mathcal{S}^{OS}$ \LyXZeroWidthSpace{} and $\mathcal{S}^{Max}$,
though the gap in performance has narrowed.
\item Downside Risk: Represented by CVaR95 and CVaR99, downside risk decreases
with $\mathcal{S}^{I+S}$ and $\mathcal{S}^{I+S,M}$ as $\omega^{i,e}$
values are reduced, whereas it remains largely unchanged for $\mathcal{S}^{OS}$
\LyXZeroWidthSpace{} and $\mathcal{S}^{Max}$.
\end{itemize}

\subparagraph{Martingale Condition for Hedge Assets:}

When comparing the Hedge PnL Mean between the original result and
the cross-parameter validation result, the latter shows a negative
sign in all test cases. Here, looking at $\mathcal{S}^{OS}$ and $\mathcal{S}^{Max}$,
there is little impact on the Model Value across a wide range of $\alpha$,
suggesting that these strategies are robust against slight deviations
from arbitrage-free conditions in scenarios caused by numerical errors.

\begin{table} \caption{Model Values and Model Risk Metrics under low vol-vol and reversed rate-vol correlation(unit: basis points)} \label{tab:pv_low_volvol} \begin{tabular}{p{1.5cm}p{1.5cm}p{1.5cm}p{1.5cm}p{1.5cm}p{1.5cm}p{1.5cm}} \toprule  &  & Model Value & NonArb Value & Hedge PnL Mean & Model Switch Value & NonArb Switch Value \\ Strategy & $\alpha$ &  &  &  &  &  \\ \midrule \multirow{4}{*}{$\mathcal{S}^{I}$} & 0.2 & 387.4 & 456.1 & -2.2 & -9.7 & 59.1 \\  & 0.4 & 344.0 & 455.7 & -3.4 & -53.0 & 58.7 \\  & 0.6 & 299.0 & 454.4 & -1.7 & -98.1 & 57.4 \\  & 0.8 & 240.8 & 453.9 & -1.8 & -156.3 & 56.8 \\ \cline{1-7} \multirow{4}{*}{$\mathcal{S}^{Max}$} & 0.2 & 426.5 & 456.9 & -4.5 & 29.4 & 59.9 \\  & 0.4 & 413.3 & 456.9 & -4.5 & 16.3 & 59.9 \\  & 0.6 & 403.8 & 456.9 & -4.5 & 6.8 & 59.9 \\  & 0.8 & 397.2 & 456.9 & -4.5 & 0.1 & 59.9 \\ \cline{1-7} \multirow{4}{*}{$\mathcal{S}^{OS}$} & 0.2 & 433.3 & 456.3 & -2.8 & 36.2 & 59.2 \\  & 0.4 & 422.4 & 456.9 & -3.6 & 25.4 & 59.9 \\  & 0.6 & 412.5 & 457.2 & -4.2 & 15.5 & 60.1 \\  & 0.8 & 405.4 & 456.7 & -4.1 & 8.4 & 59.7 \\ \cline{1-7} \multirow{4}{*}{$\mathcal{S}^{I+S}$} & 0.2 & 437.1 & 455.5 & -3.2 & 40.0 & 58.5 \\  & 0.4 & 431.0 & 456.0 & -2.1 & 34.0 & 59.0 \\  & 0.6 & 419.6 & 456.8 & -2.4 & 22.6 & 59.8 \\  & 0.8 & 406.6 & 456.8 & -1.8 & 9.6 & 59.8 \\ \cline{1-7} \multirow{4}{*}{$\mathcal{S}^{I+S,M}$} & 0.2 & 436.5 & 456.6 & -4.9 & 39.4 & 59.5 \\  & 0.4 & 432.0 & 457.0 & -3.2 & 34.9 & 59.9 \\  & 0.6 & 422.3 & 456.7 & -0.9 & 25.2 & 59.6 \\  & 0.8 & 410.2 & 457.1 & -0.4 & 13.2 & 60.1 \\ \cline{1-7} \bottomrule \end{tabular} \end{table} 

\begin{table} \caption{Distribution Metrics for Initial Model Value and Realized PnL Differences under low vol-vol and reversed rate-vol correlation(unit: basis points, excluding Loss Prob)} \label{tab:dist_low_volvol} \begin{tabular}{p{1.2cm}p{1.2cm}p{1.2cm}p{1.2cm}p{1.2cm}p{1.2cm}p{1.2cm}p{1.2cm}p{1.2cm}p{1.2cm}} \toprule  &  & P25 & P50 & P75 & IQR & Loss Prob & Expected Loss & CVaR95 & CVaR99 \\ Strategy & $\alpha$ &  &  &  &  &  &  &  &  \\ \midrule \multirow{4}{*}{$\mathcal{S}^{I}$} & 0.2 & -57.0 & 49.5 & 172.8 & 229.8 & 0.377 & 104.7 & 256.6 & 347.9 \\  & 0.4 & -8.6 & 88.9 & 202.8 & 211.4 & 0.269 & 78.8 & 184.8 & 272.2 \\  & 0.6 & 36.2 & 133.9 & 251.8 & 215.6 & 0.175 & 64.5 & 136.5 & 222.7 \\  & 0.8 & 91.5 & 189.9 & 311.7 & 220.2 & 0.081 & 54.0 & 79.0 & 162.7 \\ \cline{1-10} \multirow{4}{*}{$\mathcal{S}^{Max}$} & 0.2 & -20.6 & 5.7 & 49.9 & 70.5 & 0.454 & 20.4 & 29.4 & 29.4 \\  & 0.4 & -7.5 & 18.8 & 63.0 & 70.5 & 0.331 & 12.2 & 16.3 & 16.3 \\  & 0.6 & 2.0 & 28.3 & 72.5 & 70.5 & 0.232 & 6.0 & 6.8 & 6.8 \\  & 0.8 & 8.7 & 35.0 & 79.2 & 70.5 & 0.185 & 0.1 & 0.1 & 0.1 \\ \cline{1-10} \multirow{4}{*}{$\mathcal{S}^{OS}$} & 0.2 & -11.9 & 8.6 & 36.5 & 48.4 & 0.384 & 20.2 & 42.5 & 47.6 \\  & 0.4 & -0.9 & 13.6 & 47.8 & 48.7 & 0.270 & 12.4 & 26.2 & 31.0 \\  & 0.6 & 3.2 & 19.8 & 60.1 & 56.9 & 0.176 & 6.5 & 13.4 & 18.0 \\  & 0.8 & 8.0 & 25.8 & 67.6 & 59.6 & 0.081 & 4.7 & 6.8 & 11.6 \\ \cline{1-10} \multirow{4}{*}{$\mathcal{S}^{I+S}$} & 0.2 & -7.4 & 20.1 & 44.0 & 51.4 & 0.308 & 40.0 & 120.5 & 193.5 \\  & 0.4 & 2.8 & 21.8 & 41.6 & 38.8 & 0.222 & 26.2 & 68.9 & 114.4 \\  & 0.6 & 10.0 & 24.6 & 45.7 & 35.7 & 0.135 & 20.6 & 42.3 & 74.3 \\  & 0.8 & 10.9 & 24.2 & 56.0 & 45.1 & 0.068 & 8.8 & 11.6 & 28.9 \\ \cline{1-10} \multirow{4}{*}{$\mathcal{S}^{I+S,M}$} & 0.2 & -6.6 & 20.7 & 45.2 & 51.8 & 0.300 & 42.8 & 126.3 & 197.2 \\  & 0.4 & 3.3 & 20.4 & 38.8 & 35.5 & 0.219 & 25.0 & 65.9 & 111.4 \\  & 0.6 & 8.2 & 21.3 & 42.0 & 33.8 & 0.146 & 15.6 & 33.9 & 64.0 \\  & 0.8 & 8.3 & 24.4 & 57.7 & 49.4 & 0.069 & 6.2 & 8.4 & 20.4 \\ \cline{1-10} \bottomrule \end{tabular} \end{table}

\section{Conclusion and Future Research Directions}

In this paper, we demonstrate how applying the deep hedging approach
to the management of Bermudan swaptions can address practical challenges,
including negative P\&L trends due to improper model parameters and
downside risk with swaption hedging, and provides an enhanced management
framework. Additionally, we propose practical solutions to challenges
in applying the deep hedging approach to Bermudan swaptions, establishing
a viable framework for real-world applications. This approach is expected
to improve the overall position management environment for Bermudan
swaptions.

As for future research directions, advancing market scenario generation
models presents a promising area for further study:
\begin{enumerate}
\item Improving Hedging Model Performance with Realistic Market Scenarios:
Increasing the realism of market scenarios to better reflect actual
dynamics is essential for enhancing hedging model performance. While
the SMBM demonstrates flexibility and robustness to parameter changes
in our numerical examples, it remains a relatively simple parametric
model. More sophisticated models could improve the accuracy and adaptability
of hedging strategies.
\item Expanding the Range of Hedge assets: In this study, we limited the
hedge assets to coterminal swaptions. However, extending this methodology
to more complex trades, such as callable inverse floaters or callable
spread options, will require a broader range of swaptions as hedge
assets. Additionally, using a wider variety of swaptions could enhance
hedging efficiency for Bermudan swaptions. This expansion would likely
require more advanced scenario generation methods. Drift-removal techniques
to eliminate statistical arbitrage, using static arbitrage-free scenario
sets, could be especially beneficial for such advanced scenario generation
\cite{DHRemoveDrift}.
\end{enumerate}

\section{Disclaimer}

The opinions expressed in this article are the author\textquoteright s
own and do not reflect the view of Nomura Securities Co., Ltd. All
errors are the author\textquoteright s responsibility.

 \captionsetup[table]{labelformat=empty, font=bf, textfont=large} \begin{table}[ht] \centering \caption{Appendix: Local Correlation Matrices}
\begin{flushleft} \textbf{Correlation Matrix for $0 \leq t < 4$} \end{flushleft} \begin{flushleft} \begin{tabular}{lrrrrrrrrrr} \hline  & $S^{1,e}$ & $S^{2,e}$ & $S^{3,e}$ & $S^{4,e}$ & $S^{5,e}$ & $X^{1,e}$ & $X^{2,e}$ & $X^{3,e}$ & $X^{4,e}$ & $X^{5,e}$ \\ \hline $S^{1,e}$ & 1.000 & 0.992 & 0.964 & 0.923 & 0.878 & -0.114 & -0.074 & -0.014 & 0.005 & 0.017 \\ $S^{2,e}$ & 0.992 & 1.000 & 0.986 & 0.952 & 0.910 & -0.101 & -0.062 & -0.001 & 0.021 & 0.035 \\ $S^{3,e}$ & 0.964 & 0.986 & 1.000 & 0.986 & 0.957 & -0.083 & -0.049 & 0.012 & 0.038 & 0.053 \\ $S^{4,e}$ & 0.923 & 0.952 & 0.986 & 1.000 & 0.991 & -0.073 & -0.040 & 0.020 & 0.049 & 0.067 \\ $S^{5,e}$ & 0.878 & 0.910 & 0.957 & 0.991 & 1.000 & -0.068 & -0.035 & 0.023 & 0.056 & 0.077 \\ $X^{1,e}$ & -0.114 & -0.101 & -0.083 & -0.073 & -0.068 & 1.000 & 0.987 & 0.954 & 0.912 & 0.862 \\ $X^{2,e}$ & -0.074 & -0.062 & -0.049 & -0.040 & -0.035 & 0.987 & 1.000 & 0.986 & 0.951 & 0.907 \\ $X^{3,e}$ & -0.014 & -0.001 & 0.012 & 0.020 & 0.023 & 0.954 & 0.986 & 1.000 & 0.983 & 0.949 \\ $X^{4,e}$ & 0.005 & 0.021 & 0.038 & 0.049 & 0.056 & 0.912 & 0.951 & 0.983 & 1.000 & 0.988 \\ $X^{5,e}$ & 0.017 & 0.035 & 0.053 & 0.067 & 0.077 & 0.862 & 0.907 & 0.949 & 0.988 & 1.000 \\ \hline \end{tabular}
\end{flushleft} \vspace{1em}
\begin{flushleft} \textbf{Correlation Matrix for $4 \leq t < 5$} \end{flushleft} \begin{flushleft} \begin{tabular}{lrrrrrrrr} \hline  & $S^{2,e}$ & $S^{3,e}$ & $S^{4,e}$ & $S^{5,e}$ & $X^{2,e}$ & $X^{3,e}$ & $X^{4,e}$ & $X^{5,e}$ \\ \hline $S^{2,e}$ & 1.000 & 0.990 & 0.956 & 0.910 & -0.147 & -0.105 & -0.033 & -0.017 \\ $S^{3,e}$ & 0.990 & 1.000 & 0.983 & 0.943 & -0.134 & -0.094 & -0.021 & -0.002 \\ $S^{4,e}$ & 0.956 & 0.983 & 1.000 & 0.982 & -0.115 & -0.080 & -0.007 & 0.016 \\ $S^{5,e}$ & 0.910 & 0.943 & 0.982 & 1.000 & -0.106 & -0.073 & -0.002 & 0.026 \\ $X^{2,e}$ & -0.147 & -0.134 & -0.115 & -0.106 & 1.000 & 0.982 & 0.938 & 0.891 \\ $X^{3,e}$ & -0.105 & -0.094 & -0.080 & -0.073 & 0.982 & 1.000 & 0.981 & 0.939 \\ $X^{4,e}$ & -0.033 & -0.021 & -0.007 & -0.002 & 0.938 & 0.981 & 1.000 & 0.977 \\ $X^{5,e}$ & -0.017 & -0.002 & 0.016 & 0.026 & 0.891 & 0.939 & 0.977 & 1.000 \\ \hline \end{tabular}
\end{flushleft} \vspace{1em}
\begin{flushleft} \textbf{Correlation Matrix for $5 \leq t < 6$} \end{flushleft} \begin{flushleft} \begin{tabular}{lrrrrrr} \hline  & $S^{3,e}$ & $S^{4,e}$ & $S^{5,e}$ & $X^{3,e}$ & $X^{4,e}$ & $X^{5,e}$ \\ \hline $S^{3,e}$ & 1.000 & 0.985 & 0.939 & -0.187 & -0.144 & -0.050 \\ $S^{4,e}$ & 0.985 & 1.000 & 0.975 & -0.174 & -0.134 & -0.038 \\ $S^{5,e}$ & 0.939 & 0.975 & 1.000 & -0.150 & -0.121 & -0.027 \\ $X^{3,e}$ & -0.187 & -0.174 & -0.150 & 1.000 & 0.973 & 0.902 \\ $X^{4,e}$ & -0.144 & -0.134 & -0.121 & 0.973 & 1.000 & 0.966 \\ $X^{5,e}$ & -0.050 & -0.038 & -0.027 & 0.902 & 0.966 & 1.000 \\ \hline \end{tabular}
\end{flushleft} \vspace{1em}
\begin{flushleft} \textbf{Correlation Matrix for $6 \leq t < 7$} \end{flushleft} \begin{flushleft} \begin{tabular}{lrrrr} \hline  & $S^{4,e}$ & $S^{5,e}$ & $X^{4,e}$ & $X^{5,e}$ \\ \hline $S^{4,e}$ & 1.000 & 0.974 & -0.250 & -0.221 \\ $S^{5,e}$ & 0.974 & 1.000 & -0.240 & -0.215 \\ $X^{4,e}$ & -0.250 & -0.240 & 1.000 & 0.956 \\ $X^{5,e}$ & -0.221 & -0.215 & 0.956 & 1.000 \\ \hline \end{tabular}
\end{flushleft} \vspace{1em}
\begin{flushleft} \textbf{Correlation Matrix for $7 \leq t < 8$} \end{flushleft} \begin{flushleft} \begin{tabular}{lrr} \hline  & $S^{5,e}$ & $X^{5,e}$ \\ \hline $S^{5,e}$ & 1.000 & -0.259 \\ $X^{5,e}$ & -0.259 & 1.000 \\ \hline \end{tabular}
\end{flushleft} \vspace{1em} \end{table}
\end{document}